\documentclass[twocolumn]{aastex631}

\usepackage{verbatim}
\usepackage{graphicx}
\usepackage{appendix}
\usepackage{hyperref}

\newcommand{\msun}{M$_{\odot}$}

\newcommand{\mum}{$\mu$m}

\newcommand{\mstar}{$M_{\star}$}
\newcommand{\lstar}{$L_{\rm bol}$}

\begin{document}

\title{Older Ages for 23 Pre-Main Sequence Stars in Upper Scorpius Using Dynamical Mass-Constrained Stellar Evolutionary Models}

\author{A.~P.~M. Towner}
\affiliation{University of Arizona Department of Astronomy and Steward Observatory, 933 North Cherry Ave., Tucson, AZ 85721, USA}

\author{J.~A. Eisner}
\affiliation{University of Arizona Department of Astronomy and Steward Observatory, 933 North Cherry Ave., Tucson, AZ 85721, USA}

\author{P.~D. Sheehan}
\affiliation{National Radio Astronomy Observatory, 520 Edgemont Rd., Charlottesville, VA 22903, USA}

\author{L.~A. Hillenbrand}
\affiliation{Department of Astronomy, MC 249-17, California Institute of Technology, Pasadena, CA 91125, USA}

\author{Y.-L. Wu}
\affiliation{Department of Physics, National Taiwan Normal University, Taipei 116, Taiwan}

\begin{abstract}
We present revised stellar ages for 23 pre-main sequence K- and M-type stars in the Upper Scorpius star-forming region, derived by using stellar dynamical masses to constrain isochronal ages from five pre-main sequence stellar evolutionary models.
We find that mass-constrained stellar ages for all model sets are more consistent with the older, $\sim$8-11~Myr age for Upper Sco derived using earlier-type stars.
Additionally, applying the independent mass constraint to isochronal ages tends to 1) increase stellar ages for most model sets, and 2) decrease age scatter for individual sources between model sets.
Models that account for global magnetic fields consistently produce the best match to our observations: they change comparatively little when the mass constraint is applied, and produce 9-10~Myr ages under both unconstrained and mass-constrained conditions.
Most standard (nonmagnetic) models produce younger ages (3-5~Myr) when unconstrained, but older ages (6-9~Myr) when constrained by dynamical mass.
Our results are consistent with literature findings that suggest median disk lifetimes may be $\gtrsim$2$\times$ longer than previously thought.
\end{abstract}

\section{Introduction}
Many empirical properties and fundamental parameters of pre-main sequence (PMS) stars are expected or observed to vary as those stars evolve toward the main sequence: optical and near-infrared colors and spectral absorption features \citep{Somers2015,PerezPaolino2024}, inferred temperatures, luminosities, masses, and radii \citep{Baraffe2015,Feiden2016}, and lithium depletion rate and rotation rate \citep{Soderblom2014}, to name a few.
Understanding how these properties evolve requires accurately constraining the ages of both individual stars and overall stellar populations.
Stellar ages also impact calculations of disk lifetimes and disk evolution, including the timescales available for dust grain growth and planetesimal formation and mass accretion \citep{Ribas2014,Manara2023_review,Ben2025}.

Numerous methods exist for deriving the ages of PMS stars, including the lithium depletion boundary (LDB), absolute lithium abundances, stellar rotation, and dynamical traceback of cluster members \citep[see][and references therein]{Soderblom2014}.
Perhaps the most common method of determining age is to use stellar temperature and luminosity in combination with models of PMS stellar evolution, i.e. isochronal fitting  \citep{Cohen1979,Baraffe1998,Siess2000,Bressan2012,Baraffe2015,Feiden2016}. 
However, isochronal methods rely on accurately placing the isochrones themselves in mass-radius space; different model sets make different predictions for which $T_{\rm eff}$ and \lstar\/ correspond to a given stellar mass, radius, and age \citep[e.g. compare the magnetic and nonmagnetic predictions of][]{Feiden2016}. 
Furthermore, difficulties in accurately determining stellar properties 
have been hypothesized to contribute to observed order-of-magnitude spreads in stellar age within even a single population \citep{Soderblom2014,Somers2015,David2019}.
Young stars are also more likely to be variable, and so populations containing young stars may exhibit an intrinsic spread in \lstar\/ at a given mass and age \citep[][and references therein]{Soderblom2014}.
Additionally, a number of more recent studies have found that isochronal methods consistently return younger ages for low-mass stars than they do for higher-mass members in the same group \citep{Bell2012,Herczeg2015,Pecaut2016}. 

All these factors together suggest a need for independent calibration of PMS isochrones. 
Various studies have used independent measurements of stellar mass (or mass and radius) to perform such calibrations, typically with sample sizes of a few to a few tens of stars \citep{Rizzuto2016,David2019,Simon2019,Braun2021,Towner2025}.
These studies have used Keplerian rotation in protoplanetary disks or eclipsing binaries to derive stellar mass (or mass$+$radius). 
\citep{Soderblom2014}.

These dynamical-mass studies have found isochrones that do not include the effects of magnetic fields tend to underestimate stellar mass by $\sim$20-40\% compared to dynamical values \citep{David2019,Simon2019,Towner2025}.
Magnetic models, in contrast, tend to predict much more accurate stellar masses at the population level \citep[e.g. $<$1$\sigma$ deviation from dynamical masses;][]{Braun2021}.
Because dynamical methods do not make any direct predictions for stellar age, most studies have compared ages implied by standard models to those implied by magnetic models.
These studies have found that standard models preferentially underestimate age compared to magnetic models, though the degree of underestimation varies by sample and age: \citet{Simon2019} find a factor of 2-3$\times$ at 1-3~Myr in Taurus and Ophiuchus, \citet{Somers2015} suggest 2-10$\times$ at 3~Myr in theoretical models, and \citet{Feiden2016} and \citet{David2019} both find $\sim$1.5-2$\times$ at 5-10~Myr in Upper Sco.

A more direct comparison, however, would be between a purely isochronal age and a mass-constrained age for all PMS models.
We have recently published stellar dynamical masses for 23 K- and M-type PMS stars in the Upper Scorpius OB association \citep{Towner2025}.
These masses were derived via MCMC model fitting of Keplerian disk rotation detected in CO line emission.
In this work, we use these dynamical masses to constrain the stellar age predictions of five PMS model sets, and thus derive mass-constrained isochronal ages for each target.

The structure of the paper is as follows.
In \S~\ref{obs_section}, we describe our sample selection and target properties.
In \S~\ref{methods}, we describe our procedure for obtaining mass-constrained isochronal ages for each source.
In \S~\ref{massconstrained_analysis}, we analyze unconstrained versus mass-constrained isochronal ages for this sample, including the impacts of the mass constraint, trends with stellar mass and age, and an overall evaluation of the five model sets considered.
In \S~\ref{age_mass_literature}, we compare our findings to other dynamically-based age estimates for this region, and in \S~\ref{age_bias}, we test for age bias in our sample.
In \S~\ref{disk_lifetimes}, we consider the implications of our revised ages for protoplanetary disk lifetimes.
We summarize our results in \S~\ref{summary}.

\section{Observations}
\label{obs_section}
\subsection{The Upper Scorpius OB Association}
The Upper Scorpius OB association (hereafter Upper Sco) is the nearest massive star-forming region to Earth \citep[$\sim$145~pc,][]{Preibisch2008}.
It is one part of the larger Scorpius-Centaurus OB association \citep{Blaauw1946,Blaauw1964}. 
Although Upper Sco is older than well-known young regions such as Taurus and Lupus \citep[$\sim$1-3~Myr,][and references therein]{Andrews2013,Pascucci2016}, its precise age has been the subject of much debate for more than a decade.
Earlier estimates based on the full stellar population suggested an age of $\sim$5~Myr \citep[][and references therein]{Preibisch2002,Sartori2003,Preibisch2008}.
However, revised estimates based on only the earlier-type (B, A, F) association members suggested an age of $\sim$8-11~Myr \citep{Pecaut2012}.
Meanwhile, revised estimates based on only the later-type (K, M) members have still consistently returned an age of 5-7~Myr \citep{Herczeg2015}.

In addition to its longstanding age discrepancy, Upper Sco contains a large population of disk-bearing young stars \citep{Luhman2022,Fang2025}. 
Of the G-, K-, and M-type stars in Upper Sco, $\gtrsim$200 have dust disks detectable in the continuum \citep{Carpenter2025}, and several tens have gas disks still detectable in CO \citep{Carpenter2025,Towner2025}.
Its overall well-studied nature, seemingly mass-dependent inferred ages, and large population of CO-detectable disks make Upper Sco an excellent candidate for a systematic calibration of isochronal ages by dynamical methods.

\subsection{Sample \& Dynamical Masses}
Our sample consists of 23 K- and M-type pre-main sequence stars in Upper Sco with CO-detected disks, originally drawn from \citet{Barenfeld2016}. 
We conducted high-sensitivity follow-up observations of these targets in ALMA Cycle 7, in which we targeted CO J $=$ 3 $-$2 line emission and 345~GHz continuum (Project 2019.1.00493.S, PI: P. Sheehan).
Full details of our observing setup, data properties, and imaging procedure are given in \citet{Towner2025}.

Using our CO line data, combined with {\it Gaia} distances \citep{Gaia_i,Gaia_DR2}, we modeled the Keplerian motion of each disk to derive the enclosed stellar mass.
We used the open-source package {\tt pdspy} \citep{Sheehan2019}, which uses a Markov Chain Monte Carlo (MCMC) approach to derive best-fit disk parameters. 
Full details of the model and our fitting procedure are given in \citet{Towner2025}.
The stellar dynamical masses we derived in \citet{Towner2025} are listed in Table~\ref{source_properties}, along with source RA, Dec, {\it Gaia}-derived distance, $T_{\rm eff}$, \lstar, spectral type, and spatio-kinematic subgroup membership.

\begin{deluxetable*}{lcccccccc}
\tablecaption{Source Properties}
\tablecolumns{9}
\tablewidth{\textwidth}
\tablehead{\colhead{Source} & \colhead{RA} & \colhead{Dec} & \colhead{Distance$^a$} & \colhead{Mass$^b$} & \colhead{$L_{\rm bol}^b$} & \colhead{log($T_{\rm eff}$)$^c$} & \colhead{Spectral$^c$} & \colhead{USco$^d$}\\
& \colhead{($h$ $m$ $s$)} & \colhead{($^{\circ}$ $\arcmin$ $\arcsec$)} & \colhead{(pc)} & \colhead{(\msun)} & \colhead{(L$_{\odot}$)} & \colhead{(K)} &  \colhead{Type} & \colhead{Group}
}
\startdata
J15521088-2125372 & 15:52:10.88 & -21:25:37.20 & 168$_{-7}^{+8}$ & 0.16$^{+0.04}_{-0.03}$ & 0.010$^{+0.005}_{-0.005}$ & 3.51 (0.02) & M4 & $\delta$ Sco\\
J15530132-2114135 & 15:53:01.32 & -21:14:13.5 & 146$_{-2}^{+3}$ & 0.25$^{+1.0}_{-0.09}$ & 0.04$^{+0.02}_{-0.02}$ & 3.51 (0.02) & M4 & $\delta$ Sco\\
J15534211-2049282$^e$ & 15:53:42.11 & -20:49:28.2 & 136$_{-3}^{+3}$ & 0.6$^{+0.6}_{-0.1}$ & 0.07$^{+0.03}_{-0.03}$ & 3.52 (0.02) & M3.5 & Ant\\
J15562477-2225552 & 15:56:24.77 & -22:25:55.2 & 141$_{-2}^{+2}$ & $\leq$1.45 & 0.05$^{+0.02}_{-0.02}$ & 3.51 (0.02) & M4 & $\delta$ Sco\\
J16001844-2230114$^e$ & 16:00:18.44 & -22:30:11.4 & 138$_{-8}^{+9}$ & 0.5$^{+3.0}_{-0.3}$ & 0.05$^{+0.02}_{-0.02}$ & 3.50 (0.02) & M4.5 & \nodata\\
J16014086-2258103$^e$ & 16:01:40.85 & -22:58:11.3 & 124$_{-2}^{+2}$ & 0.25$^{+0.25}_{-0.05}$ & 0.07$^{+0.03}_{-0.03}$ & 3.51 (0.02) & M4 & $\delta$ Sco\\
J16020757-2257467 & 16:02:07.57 & -22:57:46.7 & 140$_{-1}^{+1}$ & 0.54$^{+0.05}_{-0.04}$ & 0.15$^{+0.06}_{-0.07}$ & 3.54 (0.02) & M2.5 & $\delta$ Sco\\
J16035767-2031055 & 16:03:57.67 & -20:31:05.5 & 142.6$_{-0.8}^{+0.8}$ & 4.0$^{+4.0}_{-2.0}$ & 0.6$^{+0.3}_{-0.3}$ & 3.64 (0.01) & K5 & $\delta$ Sco\\
J16035793-1942108 & 16:03:57.93 & -19:42:10.8 & 158$_{-2}^{+2}$ & 0.56$^{+0.05}_{-0.04}$ & 0.13$^{+0.05}_{-0.06}$ & 3.55 (0.02) & M2 & $\beta$ Sco\\
J16062277-2011243 & 16:06:22.77 & -20:11:24.3 & 151$_{-2}^{+2}$ & 0.3$^{+0.3}_{-0.1}$ & 0.05$^{+0.02}_{-0.02}$ & 3.49 (0.02) & M5 & $\beta$ Sco\\
J16075796-2040087$^e$ & 16:07:57.96 & -20:40:08.7 & 159$_{-6}^{+7}$ & 1.9$^{+0.2}_{-0.2}$ & 0.09$^{+0.04}_{-0.05}$ & 3.57 (0.02) & M1 & $\nu$ Sco\\
J16081566-2222199 & 16:08:15.66 & -22:22:19.9 & 140$_{-1}^{+2}$ & 0.5$^{+0.1}_{-0.08}$ & 0.14$^{+0.06}_{-0.06}$ & 3.53 (0.02) & M3.25 & $\delta$ Sco\\
J16082324-1930009 & 16:08:23.24 & -19:30:00.9 & 138$_{-1}^{+1}$ & 0.91$^{+0.04}_{-0.02}$ & 0.2$^{+0.1}_{-0.1}$ & 3.59 (0.01) & K9 & $\delta$ Sco\\
J16090075-1908526 & 16:09:00.75 & -19:08:52.6 & 138$_{-1}^{+1}$ & 0.71$^{+0.03}_{-0.03}$ & 0.3$^{+0.1}_{-0.1}$ & 3.59 (0.01) & K9 & $\nu$ Sco\\
J16095933-1800090 & 16:09:59.33 & -18:00:09.0 & 136$_{-2}^{+2}$ & 0.22$^{+0.05}_{-0.02}$ & 0.07$^{+0.04}_{-0.03}$ & 3.51 (0.02) & M4 & $\nu$ Sco\\
J16104636-1840598 & 16:10:46.36 & -18:40:59.8 & 143$_{-3}^{+3}$ & 0.13$^{+0.07}_{-0.05}$ & 0.03$^{+0.01}_{-0.02}$ & 3.50 (0.02) & M4.5 & $\nu$ Sco\\
J16113134-1838259 B$^{e,f}$ & 16:11:31.31 & -18:38:27.7 & 157$_{-4}^{+5}$ & 1.53$^{+0.14}_{-0.12}$ & 0.4$^{+0.2}_{-0.1}$ & 3.60 (0.01) & K7 & \nodata \\
J16115091-2012098 & 16:11:50.91 & -20:12:09.8 & 152$_{-4}^{+4}$ & 0.31$^{+0.05}_{-0.03}$ & 0.10$^{+0.04}_{-0.04}$ & 3.52 (0.02) & M3.5 & $\delta$ Sco\\
J16123916-1859284 & 16:12:39.16 & -18:59:28.4 & 139$_{-2}^{+2}$ & 0.76$^{+0.11}_{-0.07}$ & 0.3$^{+0.1}_{-0.1}$ & 3.58 (0.01) & M0.5 & $\nu$ Sco\\
J16142029-1906481 & 16:14:20.29 & -19:06:48.1 & 143$_{-2}^{+3}$ & 1.32$^{+0.06}_{-0.06}$ & 0.05$^{+0.02}_{-0.02}$ & 3.59 (0.01) & M0 & $\nu$ Sco\\
J16143367-1900133 & 16:14:33.61 & -19:00:14.8 & 142$_{-2}^{+2}$ & 0.23$^{+0.05}_{-0.04}$ & 0.10$^{+0.04}_{-0.04}$ & 3.53 (0.02) & M3 & $\nu$ Sco\\
J16163345-2521505 & 16:16:33.45 & -25:21:50.5 & 163$_{-1}^{+1}$ & 0.8$^{+0.1}_{-0.1}$ & 0.17$^{+0.08}_{-0.07}$ & 3.58 (0.01) & M0.5 & $\sigma$ Sco\\
J16181904-2028479 & 16:18:19.04 & -20:28:47.9 & 138$_{-2}^{+2}$ & 0.25$^{+0.15}_{-0.05}$ & 0.03$^{+0.02}_{-0.02}$ & 3.50 (0.02) & M4.75 & $\nu$ Sco
\enddata
\tablenotetext{a}{Distances are derived from {\it Gaia} parallax measurements.}
\tablenotetext{b}{Dynamical masses for all sources come from \citet{Towner2025}, and are derived via modeling of the Keplerian disk rotation in CO. Luminosities come from \citet{Towner2025}, and are derived using OIR photometry and BT Settl models for all sources except J16113134-1838259 B; that source comes from  \citet{Eisner2005}, who use OIR spectroscopy and SED modeling to derive \lstar\/ and $T_{\rm eff}$.}
\tablenotetext{c}{The spectral type and effective temperature for J16113134-1838259 B come from the spectroscopic analysis of \citet{Eisner2005}. For all other sources, spectral types are taken from \citet{Luhman2012}, which collates spectral types from spectroscopic studies in the literature. For these sources, effective temperatures are taken from \citet{Barenfeld2016}, who derive $T_{\rm eff}$ using the \citet{Luhman2012} spectral types and the temperature scales of \citet{Schmidt-Kaler1982}, \citet{Straizys1992}, and \citet{Luhman1999}, assuming an uncertainty of $\pm$1 spectral subclass.}
\tablenotetext{d}{These are subgroup(s) within the larger Sco-Cen complex with which each source is most consistent, using the system of \citet{Ratzenbock2023a}. 
Sources with a ``\nodata'' are not included in that literature sample. 
Source designations indicate subgroups within Upper Sco specifically (``$\nu$,'' ``$\delta$,'' etc).
The subgroup Antares (``Ant'') lies within Upper Sco.}
\tablenotetext{e}{These sources are either known binaries or have a candidate companion within the 1.3~mm continuum disk emission, as reported in  \citet{Barenfeld2016} and \citet{Barenfeld2017_disksize}.}
\tablenotetext{f}{Source J16113134-1838259 B is a spectroscopic binary. The \lstar\/ and $T_{\rm eff}$ reported above are for the higher-mass member. The lower-mass companion has log($T_{\rm eff}$) $=$ 3.58$\pm$0.01 and log($L_{bol}$) $=$ $-$0.59$\pm$0.15, as reported in \citet{Eisner2005}.}
\label{source_properties}
\end{deluxetable*}

\section{Methods \& Results}
\label{methods}
Keplerian fitting of the protostellar disk does not directly yield a stellar age.
Therefore, following methods established in the literature \citep[e.g.][]{Soderblom2014,Rizzuto2016,David2019}, we use  five pre-main sequence evolutionary tracks to derive two stellar ages for each source: one age that depends only on \lstar, $T_{\rm eff}$, and the PMS tracks, and one age that depends on those factors and also on the stellar dynamical mass ($M_{\rm dyn}$).
For this work, we use the same PMS model sets as in \citet{Towner2025}: the BHAC15 tracks of \citet{Baraffe2015}, the PARSEC v1.1 and 1.2S tracks of \citet{Bressan2012} and \citet{Chen2014}, respectively, and both the non-magnetic and magnetic tracks of \citet{Feiden2016}.

Our procedure is as follows: for each source, we generate Gaussian distributions in stellar temperature and luminosity using the Python package {\tt numpy.random.normal}.
The distributions are centered on the $T_{\rm eff}$ and \lstar\/ values reported in Table~\ref{source_properties}, with widths corresponding to the respective $T_{\rm eff}$ and \lstar\/ uncertainties.
Each distribution has 10,000 points.
We feed these $T_{\rm eff}$ and \lstar\/ into each set of evolutionary tracks using linear interpolation between the grid points, and obtain corresponding distributions of isochronal \mstar\/ and $\tau$.
We take the median of the \mstar\/ and $\tau$ distributions as the isochronal masses and ages, respectively. 
We take as our uncertainties the scaled median absolute deviation (1.4826$\times$MAD)\footnote{For a normal distribution, 1.4826$\times$MAD is equivalent to 1$\sigma$.} of the \mstar\/ and $\tau$ distributions.
We refer to this $\tau$ value $-$ which depends only on $T_{\rm eff}$, \lstar, and the isochrones $-$ as the ``unconstrained isochronal age,'' or $\tau_{\rm all}$.

Next, we discard all (\mstar$,\tau$) pairs for which the isochronal \mstar\/ does not agree with our dynamical mass within uncertainties (see Table~\ref{source_properties}).
We then calculate the median $\tau$ and scaled MAD of all remaining ages.
This gives us what we call the ``dynamical mass-constrained age,'' or $\tau_{\rm dyn}$.
This age depends on our dynamical mass, in addition to \lstar, $T_{\rm eff}$, and the isochrones.

\begin{figure*}
    \centering
    \includegraphics[width=\textwidth]{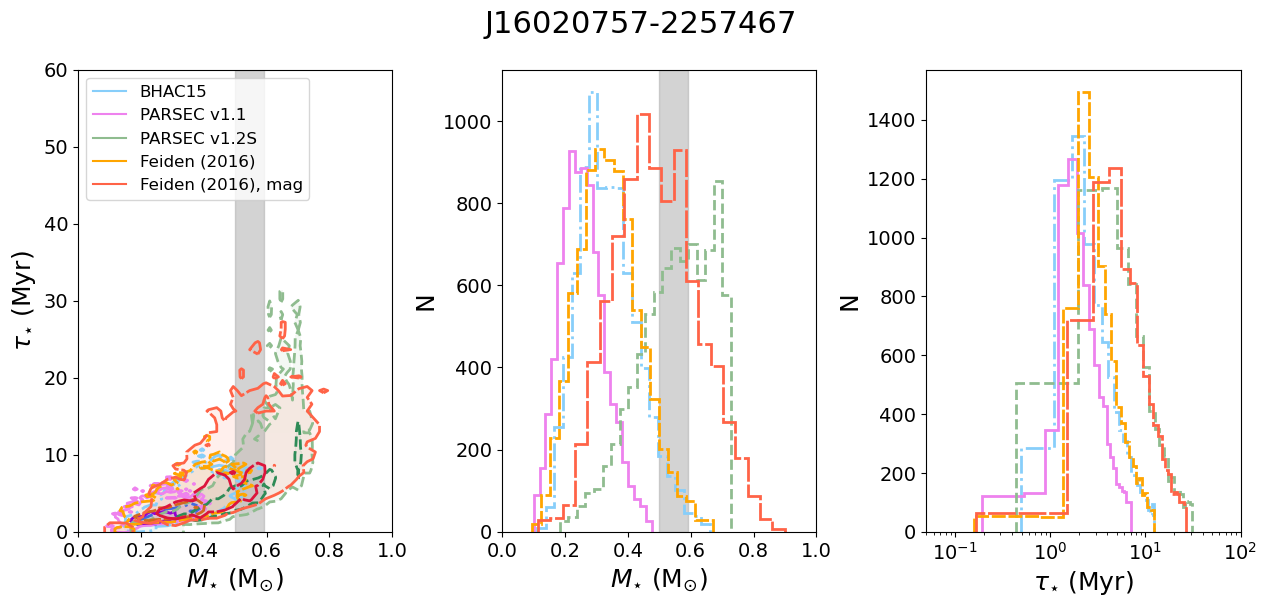}
    \caption{Isochronal masses and ages for example source J16020757$-$2257467. {\it Left:} Contour plots of isochronal age ($\tau_{\rm all}$) versus isochronal mass ($M_{iso}$) for all five model sets considered in this work. The vertical gray bar denotes the disk-derived dynamical mass, including uncertainties. The portions of each set of contours that overlap the gray bar are what comprise our mass-constrained age ($\tau_{\rm dyn}$) for a given source. {\it Center:} Histograms of the distribution of isochronal mass for each of the five model sets. The gray bar again denotes the derived dynamical mass, including uncertainties. {\it Right:} Histograms of the distribution of unconstrained isochronal age ($\tau_{\rm all}$) for each model set. Note the logarithmic scale on the x-axis. The mass-constrained ages for this source ($\tau_{\rm dyn}$) are: BHAC15: 11$\pm$8~Myr, PARSEC v1.1: 8$\pm$5~Myr, PARSEC v1.2S: 6$\pm$4~Myr, nonmagnetic \citet{Feiden2016}: 10$\pm$5~Myr, magnetic \citet{Feiden2016}: 9$\pm$5~Myr.
    }
    \label{mass_histogram}
\end{figure*}

Figure~\ref{mass_histogram} shows a visual representation of this process.
In the left-hand panel is a contour plot of the (\mstar$,\tau$) distributions for all five model sets for source J16020757$-$2257467.
The middle and right-hand panels show histograms of the distributions of \mstar\/ and $\tau$.
The dynamical mass range is shown as a gray bar. 
Our process of applying a dynamical-mass constraint to the stellar ages means that, instead of using the full distribution of ages, we use only those ages for each model set that fall within the gray bar.

We report all ages in Table~\ref{ages_table}.
Columns 2-6 list the median and scaled MAD of the unconstrained isochronal $\tau$ for each source and model set. 
Columns 7-11 list the median and scaled MAD of the mass-constrained $\tau$ for each source and model set.
Entries with a ``\nodata'' indicate that there was no overlap within uncertainties between the dynamical and isochronal masses for that source$+$model set.

Overall, we find that stellar ages are highly model-dependent, consistent with previous literature findings \citep[e.g.][]{David2019,Ratzenbock2023b}.
The BHAC15, PARSEC v1.1, and non-magnetic \citet{Feiden2016} models tend to produce lower ages than the PARSEC v1.2S and magnetic \citet{Feiden2016} models in the unconstrained condition.
Both the PARSEC v1.2S and \citet{Feiden2016} magnetic models tend to produce older ages than the other three models under both the unconstrained and mass-constrained conditions.

\rotate
\begin{deluxetable*}{l|ccccc|ccccc}
\tablecaption{Stellar Ages By PMS Evolutionary Model Set}
\tablecolumns{11}
\tablewidth{\textwidth}
\tablehead{
\colhead{} & \multicolumn{5}{c}{All Isochronal Ages$^a$} & \multicolumn{5}{c}{Dynamical Mass-constrained Isochronal Ages$^a$} \\
\colhead{Field} & \colhead{BHAC15} & \colhead{PARSEC v1.1} & \colhead{PARSEC v1.2S} & \colhead{F16, n-m} & \colhead{F16, m} & \colhead{BHAC15} & \colhead{PARSEC v1.1} & \colhead{PARSEC v1.2S} & \colhead{F16, n-m} & \colhead{F16, m}
}
\startdata
J15521088-2125372 & 50 (51) & 18 (9) & 49.3 (0.3) & 45 (7) & 81 (81) & 37 (25) & 27 (2) & 49.0 (0.3) & 44 (8) & 42 (29) \\
J15530132-2114135 & 9 (7) & 4 (2) & 34 (22) & 8 (6) & 15 (12) & 11 (9) & 7 (3) & 34 (22) & 12 (9) & 16 (12) \\
J15534211-2049282$^b$ & 6 (5) & 3 (2) & 19 (15) & 6 (4) & 10 (8) & 28 (15) & \nodata & 25 (14) & 24 (12) & 26 (14) \\
J15562477-2225552 & 7 (5) & 4 (2) & 25 (19) & 6 (4) & 12 (8) & \nodata & \nodata & \nodata & \nodata & \nodata \\
J16001844-2230114$^b$ & 5 (3) & 3 (1) & 18 (15) & 5 (3) & 8 (5) & 7 (4) & 5 (2) & 18 (14) & 8 (5) & 9 (6) \\
J16014086-2258103$^b$ & 4 (3) & 3 (1) & 14 (12) & 4 (2) & 8 (5) & 6 (4) & 4 (2) & 12 (9) & 7 (4) & 8 (5) \\
J16020757-2257467 & 4 (3) & 3 (1) & 8 (7) & 4 (3) & 8 (6) & 11 (8) & 8 (5) & 6 (4) & 10 (5) & 9 (5) \\
J16035767-2031055 & 6 (5) & 5 (4) & 5 (4) & 6 (5) & 14 (10) & \nodata & \nodata & \nodata & \nodata & \nodata \\
J16035793-1942108 & 7 (6) & 4 (2) & 15 (12) & 7 (5) & 14 (11) & 16 (11) & 12 (5) & 11 (9) & 14 (9) & 13 (9) \\
J16062277-2011243 & 4 (2) & 3 (1) & 13 (10) & 4 (2) & 6 (3) & 8 (5) & 5 (2) & 14 (10) & 9 (5) & 9 (5) \\
J16075796-2040087$^b$ & 24 (22) & 10 (8) & 46 (5) & 24 (22) & 41 (33) & \nodata & \nodata & \nodata & \nodata & \nodata \\
J16081566-2222199 & 3 (2) & 2 (1) & 7 (6) & 3 (2) & 6 (5) & 9 (6) & 6 (3) & 7 (5) & 8 (5) & 8 (4) \\
J16082324-1930009 & 10 (8) & 6 (4) & 11 (9) & 11 (9) & 22 (15) & \nodata & \nodata & \nodata & \nodata & 13 (4) \\
J16090075-1908526 & 5 (3) & 4 (2) & 6 (3) & 5 (3) & 12 (7) & 7 (4) & 7 (3) & 6 (3) & 7 (4) & 37 (17) \\
J16095933-1800090 & 4 (3) & 3 (1) & 14 (12) & 4 (3) & 7 (5) & 5 (3) & 4 (2) & 4 (3) & 5 (3) & 6 (3) \\
J16104636-1840598 & 9 (7) & 5 (2) & 40 (14) & 9 (6) & 16 (12) & 6 (4) & 5 (2) & 49.0 (0.4) & 7 (4) & 9 (5) \\
J16113134-1838259 B$^b$ & 4 (2) & 2 (1) & 4 (2) & 3 (2) & 9 (5) & \nodata & \nodata & \nodata & \nodata & \nodata \\
J16115091-2012098 & 4 (3) & 2 (1) & 10 (9) & 4 (2) & 7 (5) & 5 (3) & 5 (2) & 4 (2) & 5 (3) & 6 (3) \\
J16123916-1859284 & 4 (2) & 3 (1) & 5 (3) & 4 (2) & 9 (5) & 9 (4) & \nodata & 6 (3) & 9 (4) & 9 (5) \\
J16142029-1906481 & 154 (90) & 28 (0) & 48.8 (0.5) & 49.6 (0.3) & 141 (40) & \nodata & \nodata & \nodata & \nodata & \nodata \\
J16143367-1900133 & 5 (4) & 3 (2) & 14 (12) & 5 (4) & 10 (7) & 3 (1) & 3 (1) & 2 (1) & 4 (2) & 5 (2) \\
J16163345-2521505 & 10 (8) & 6 (4) & 14 (11) & 11 (8) & 21 (14) & 14 (7) & 8 (2) & 12 (7) & 14 (6) & 19 (10) \\
J16181904-2028479 & 8 (6) & 4 (2) & 35 (21) & 8 (6) & 14 (11) & 16 (12) & 9 (4) & 30 (27) & 17 (12) & 18 (13) \\
\enddata
\tablenotetext{a}{Stellar ages derived using each of the 5 PMS evolutionary models. Ages in columns 2-6 are the median of all ages returned by that track within 3$\sigma$. Ages in columns 7-11 are the median age of only those tracks within 3$\sigma$ that {\it also} returned a stellar mass that agrees with the dynamical mass within uncertainties. Entries with a ``\nodata'' do not have any PMS tracks that meet the mass-constraint criterion.}
\tablenotetext{b}{These sources have confirmed or candidate companions; see Table~\ref{source_properties} for further details.}
\label{ages_table}
\end{deluxetable*}

There are five sources with no mass-constrained ages in any model set, and four additional sources with no mass-constrained age in at least one model set.
The PARSEC v1.1 models are the most likely model set to have no overlap with the dynamical masses (8/23 sources), followed by BHAC15, PARSEC v1.2S, and non-magnetic \citet{Feiden2016} models (6/23 sources), and finally the \citet{Feiden2016} magnetic models (5/23 sources).
The five sources with no mass-constrained ages at all are, without exception, those sources with \mstar\/ $\geq$ 1~\msun.
Only two of these have confirmed or candidate companions; the others simply have anomalously-high dynamical masses, as discussed in \citet{Towner2025}.
Of the sources with mass-constrained ages in some but not all model sets, there is no universal mass cutoff, though these sources do tend to have \mstar\/ $>$ 0.5~\msun.

All five model sets cover up to at least 1.4~\msun, and some reach $>$4.0~\msun\/ (the maximum dynamical mass in our sample), so this failure cannot be purely an issue of model grid coverage.
Instead, it appears to be a result of a mismatch between the derived \lstar\/ and $T_{\rm eff}$ from optical and near-infrared photometry and the CO disk-derived \mstar.
For the two sources with \mstar\/ $>$ 1~\msun\/ that have confirmed or candidate companions, this mismatch is easily understood. 
For the three sources with no companions but also no mass-constrained ages, the picture is more complicated.
All three of these sources are discussed individually in Appendix~B of \citet{Towner2025}.
In that work, we raised the possibility that the CO data for these three sources contained some components that were insufficiently captured by the Keplerian-disk model; the mismatch between the stars' \lstar-$T_{\rm eff}$ locations on the Hertzsprung-Russell diagram and their CO-derived dynamical masses is consistent with this possibility.

\section{Mass-Constrained Isochronal Ages}
\label{massconstrained_analysis}
\begin{figure*}
    \centering
    \includegraphics[width=\textwidth]{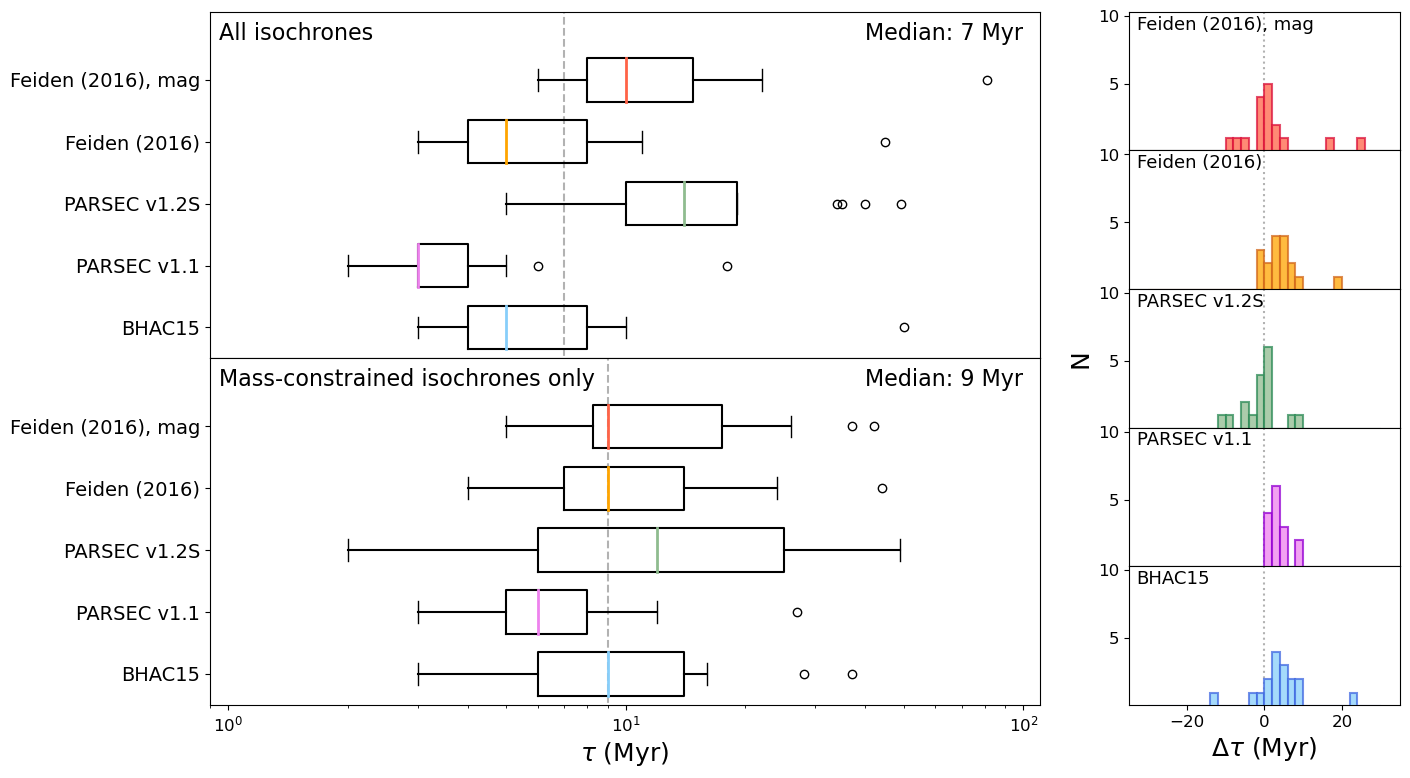}
    \caption{Box-and-whisker plots showing the distribution of stellar ages for each of the five sets of isochrones in this work. Median values are shown as colored lines in each box, and the left and right edge of each box are the first and third quartile values, respectively. The lower and upper whiskers end at the last data point to fall within Q1 - 1.5IQR and Q3+1.5IQR, respectively, where IQR is the inter-quartile range (Q3 - Q1). Fliers are shown as open circles.
    {\it Left, Top:} Distribution of ages for each model set using all ages returned by the isochrones within 3$\sigma$ (i.e. the 99.7\% percentile). The median stellar age across all model sets is shown as a dashed gray line. This panel excludes sources that do not have a corresponding mass-constrained age; see Table~\ref{ages_table} for those data. {\it Left, Bottom:} Distribution of ages for each model set using only the tracks within the 99.7\% percentile that {\it also} return stellar masses that agree with the dynamical mass within uncertainties. These are referred to as ``mass-constrained ages'' in the text. The median mass-constrained stellar age is shown as a dashed gray line. The median age across all isochrones increases from 7$\pm$4~Myr in the unconstrained case to 9$\pm$6~Myr when the mass constraint is applied.
    {\it Right Column:} Histograms showing the change in stellar age ($\Delta\tau$ $=$ $\tau_{constrained}$ - $\tau_{\rm all}$) for each source, by model set. Bins are 2~Myr wide. $\Delta\tau$ $=$ 0 is shown as a grey dotted line in all panels. The BHAC15, PARSEC v1.1, and non-magnetic \citet{Feiden2016} models tend to show an increase in age (median $\Delta\tau$ $=$ (2-3) $\pm$ (3-4)~Myr), while the magnetic \citet{Feiden2016} and PARSEC v1.2S model sets show a median of no change in age (median $\Delta\tau$ $=$ 0 $\pm$ 3~Myr for both).
    }
    \label{age_boxplots}
\end{figure*}

In Figure~\ref{age_boxplots}, we show boxplots of the distribution of ages under both the unconstrained ($\tau_{\rm all}$) and mass-constrained ($\tau_{\rm dyn}$) conditions.
This figure also shows histograms of how age changes for each source and model set when the mass constraint is applied. 
In Table~\ref{agestats}, we present age statistics (minimum, maximum, median, mean) for each model set under both the unconstrained and mass-constrained conditions.
This table also presents summary statistics for the ratio of unconstrained to mass-constrained ages ($\tau_{\rm all}$/$\tau_{\rm dyn}$), absolute changes in age ($\Delta\tau$), and intra-source age spreads ($\sigma_{\tau}$).

When we apply the dynamical mass constraint to the isochronal ages, we find two consistent effects: 1) stellar age tends to increase, and 2) the ages derived from different model sets tend to have less scatter for each individual source. 
These trends persist at both the individual level and the population level.
These results suggest two things: first, that most purely isochronal methods are systematically underestimating stellar ages for $\sim$10~Myr, K- and M-type PMS stars, and second, that applying an independent mass constraint can help relieve some of the model discordance in isochronal ages that persists in the literature.
We discuss each of these results in greater detail in the following two subsections.

Because some sources do not have mass-constrained ages in some or all model sets, we restrict our statistical analysis in the following subsections to only those sources that have {\it both} an unconstrained and a mass-constrained age for a given model set.
The boxplots in the top panel of Figure~\ref{age_boxplots} likewise include only those sources with two ages. 
However, we have calculated all statistics for the full unconstrained sample as well, and find that they are identical within uncertainties with the restricted unconstrained sample.
The statistics for our full, unconstrained sample are listed in the caption of Table~\ref{agestats}.
Finally we note that, in \citet{Towner2025}, we restricted our analysis to only those sources with M$_{\star,\,\rm dyn}$ $\leq$ 1~\msun.
Because the sources with M$_{\star,\,\rm dyn}$ $>$ 1~\msun\/ have no mass-constrained ages in Table~\ref{ages_table}, our analysis here excludes those more massive sources automatically.

\subsection{Impacts of Using Dynamical Mass to Constrain Isochronal Age}

\subsubsection{Increasing Stellar Ages}

The median unconstrained ages for the BHAC15, PARSEC v1.1, and non-magnetic \citet{Feiden2016} models are 5$\pm$1, 3$\pm$1, and 5$\pm$1~Myr, respectively. 
These values are in line with the younger ages for Upper Sco reported in e.g. \citet{Preibisch2002} and \citet{Herczeg2015} using later-type (K,M) stars.
In contrast, the PARSEC v1.2S and magnetic \citet{Feiden2016} models have median unconstrained ages of 14$\pm$7 and 10$\pm$5~Myr, respectively, which are more consistent with the comparatively older ages derived by e.g. \citet{Pecaut2012} using earlier-type (B,A,F) stars.
In other words, using unconstrained isochronal ages reproduces the age discrepancy for Upper Sco that has long been observed in the literature \citep{Preibisch2008,Pecaut2012}.

When the mass constraint is applied, stellar age shows a strong tendency to increase.
The median mass-constrained ages rise to 6$-$12~Myr, depending on the model set. 
The median BHAC15 and non-magnetic \citet{Feiden2016} ages rise to 9$\pm$4 and 9$\pm$5~Myr, respectively (a $+$4~Myr increase in both cases), i.e. they now align more closely with the higher literature ages derived using earlier-type stars.
The PARSEC v1.1 ages remain comparatively young: their median mass-constrained age is 6$\pm$3~Myr which, while it is a $+$3~Myr increase from the unconstrained age for that model set, is still more consistent with the younger literature ages for Upper Sco.
The median mass-constrained ages for the PARSEC v1.2S and magnetic \citet{Feiden2016} models are 12$\pm$9 and 9$\pm$5~Myr, respectively, i.e. they change comparatively little when the mass constraint is applied and are still consistent with the higher age estimates for Upper Sco.
We do note that, in contrast to the other model sets, the PARSEC v1.2S and \citet{Feiden2016} magnetic models show a {\it decrease} in median stellar age when the mass constraint is applied ($-$2 and $-$1~Myr, respectively).
This suggests they may be overestimating age slightly at the population level, although this picture is complicated by the statistics examining age change for individual sources (see below).

\begin{deluxetable}{lcccc}
\tablecaption{Age Statistics by Model Set and Mass Constraint}
\tablecolumns{5}
\tablewidth{\textwidth}
\tablehead{
 \colhead{Model Set} & \colhead{Min} & \colhead{Max} & \colhead{Median} & \colhead{Mean} 
}
\startdata
\cutinhead{Absolute Age (unconstrained)$^a$}
BHAC15 & 3 & 50 & 5 (1) & 8 (11)\\
PARSEC v1.1 & 2 & 18 & 3 (1) & 4 (4)\\
PARSEC v1.2S & 5 & 49.3 & 14 (7) & 19 (13)\\
Feiden (2016), non-mag & 3 & 45 & 5 (1) & 8 (9)\\
Feiden (2016), magnetic & 6 & 81 & 10 (5) & 15 (17)\\
%
\cutinhead{Absolute Age (mass-constrained)}
BHAC15 & 3 & 37 & 9 (4) & 12 (9)\\
PARSEC v1.1 & 3 & 27 & 6 (3) & 8 (6)\\
PARSEC v1.2S & 2 & 49 & 12 (9) & 17 (15)\\
Feiden (2016), non-mag & 4 & 44 & 9 (4) & 12 (9)\\
Feiden (2016), magnetic & 5 & 42 & 9 (5) & 15 (10)\\
%
\cutinhead{Age Ratio ($\tau_{i,\,\rm all}$/$\tau_{i,\,\rm dyn}$)}
BHAC15 & 0.21 & 1.67 & 0.7 (0.3) & 0.7 (0.4)\\
PARSEC v1.1 & 0.33 & 1.0 & 0.6 (0.2) & 0.6 (0.2)\\
PARSEC v1.2S & 0.76 & 7.0 & 1.0 (0.2) & 2 (2)\\
Feiden (2016), non-mag & 0.25 & 1.29 & 0.6 (0.3) & 0.7 (0.3)\\
Feiden (2016), magnetic & 0.32 & 2.0 & 1.0 (0.3) & 1.1 (0.5)\\
\cutinhead{Absolute Change in Age ($\Delta\tau$ $=$ $\tau_{i,\,\rm dyn}$ $-$ $\tau_{i,\,\rm all}$)}
BHAC15 & $-$13 & 22 & 2 (4) & 3 (7)\\
PARSEC v1.1 & 0 & 9 & 3 (3) & 3 (3)\\
PARSEC v1.2S & $-$12 & 9 & 0 (3) & -2 (5)\\
Feiden (2016), non-mag & $-$2 & 18 & 3 (3) & 4 (5)\\
Feiden (2016), magnetic & $-$39 & 25 & 0 (3) & -1 (12)\\
\cutinhead{Intra-source Age Spread ($\sigma_{\tau}$ $=$ $\tau_{i,\,\rm max}$ $-$ $\tau_{i,\,\rm min}$)}
Unconstrained ages$^b$ & 5 & 63 & 11 (6) & 17 (15) \\
Mass-constrained ages & 2 & 44 & 8 (7) & 13 (12) \\
\enddata
\tablenotetext{a}{The unconstrained age statistics shown here include only those sources which have both an unconstrained and a mass-constrained age for a given model set. Minimum, maximum, median, and mean unconstrained ages for the full sample are: BHAC15: 3, 154, 6$\pm$2, 15$\pm$32~Myr; PARSEC v1.1: 2, 28, 3.5$\pm$0.7, 6$\pm$6~Myr; PARSEC v1.2S: 4, 49.3, 14$\pm$10, 20$\pm$15~Myr; \citet{Feiden2016} non-magnetic: 3, 49.6, 6$\pm$2, 10$\pm$12~Myr; \citet{Feiden2016} magnetic: 6, 141, 11$\pm$5, 22$\pm$31~Myr.}
\tablenotetext{b}{The values shown in this row include only those sources that also have a mass-constrained age. The unconstrained $\sigma_\tau$ statistics for the full sample are: minimum $=$ 5~Myr, maximum $=$ 126~Myr, median $=$ 11$\pm$7~Myr, mean $=$ 22$\pm$26~Myr.}
\label{agestats}
\end{deluxetable}

In order to more fully quantify how the stellar ages change when constrained, we calculate for each source$+$model set the ratio of the unconstrained to mass-constrained age ($\tau_{i,\,\rm all}$/$\tau_{i,\,\rm dyn}$, for a given source$+$model set combination $i$). 
We find that the BHAC15, PARSEC v1.1, and non-magnetic \citet{Feiden2016} models have median age ratios of 0.6$-$0.7, i.e., the unconstrained ages exhibit a median underestimation of 30-40\% relative to the mass-constrained age or a factor of 1.4$\times$ to 1.67$\times$ relative to the unconstrained age.
These three models exhibit a median absolute change in age of $\Delta\tau$ $=$ 2-3~Myr, where we define $\Delta\tau_i$ $=$ $\tau_{i,\,\rm dyn}$ $-$ $\tau_{i,\,\rm all}$. 
The PARSEC v1.2S and magnetic \citet{Feiden2016} models, in contrast, both have median age ratios of 1.0 and $\Delta\tau$ $=$ 0~Myr, i.e. no change.
Their population-level age decreases (discussed above) appear to be driven by absolute age changes in a few outlier sources; their relative age changes remain fairly low. 

For all models except the \citet{Feiden2016} magnetic, the scaled MAD of the sample {\it increases} when the mass constraint is applied, i.e., the sample spans a wider range of ages. 
This cannot be a consequence of a smaller sample size in the mass-constrained ages; as noted above, we are comparing identical sources in the unconstrained and mass-constrained conditions.
Rather, this increase in scatter appears to be a real effect in the data, and can be seen in the boxplot distributions and histograms in Figure~\ref{age_boxplots}: the inter-quartile distances preferentially increase when the mass constraint is applied, and the distributions (with the exception of the magnetic models) tend to become more even overall.
Additionally, four of the model sets have at least three sources for which stellar age {\it decreases} when the mass constraint is applied.
The sources with a decrease in age tend to be consistent across model sets, rather than being randomly distributed.

We suggest that this increase in overall scatter, and consistent decrease in age for some sources, may be due to intrinsic age spread in the Upper Sco region.
\citet{Ratzenbock2023a} identify nine separate spatio-kinematic subgroups in Upper Sco, and \citet{Ratzenbock2023b} derive ages for each group and identify age gradients across the Sco-Cen complex.
Our sources span five of the nine Upper Sco subgroups ($\delta$, $\beta$, $\nu$, $\sigma$, and Antares). The age estimates in \citet{Ratzenbock2023b} for these five subgroups range from 3.9$^{+0.2}_{-0.5}$~Myr for $\nu$ Sco to 12.7$^{+0.4}_{-1.7}$~Myr for the Antares group\footnote{\citet{Ratzenbock2023b} derive four ages for each subgroup, using two different color-magnitude diagrams and the BHAC15 and PARSEC v1.2S model sets. The ages we quote here are the minimum and maximum age derived for the groups covered by our sample. The uncertainties quoted reflect the upper and lower age limits listed in Table 1 of \citet{Ratzenbock2023b}. That work uses the 1$\sigma$ highest-density interval in the marginalized Probability Density Function (PDF) to derive upper and lower age limits for each group$+$method of age derivation.}.
The spreads in mass-constrained age we observe within each model set for our sample are consistent with the spread in subgroup ages from \citet{Ratzenbock2023b}.

Finally, we examine the uncertainties on each age, including whether uncertainties tend to increase or decrease for individual sources when the mass constraint is applied.
We calculate the minimum, maximum, median, and scaled MAD of the relative uncertainty ($\tau_{\rm uncertainty}$/$\tau_{\rm dyn}$) for each source and model set under both conditions. 
Under both conditions, the PARSEC v1.1 models tend to produce the lowest relative uncertainties while the PARSEC v1.2S models tend to produce the highest. 

We find that the median relative uncertainty decreases for all five model sets when the mass constraint is applied: median uncertainties for the unconstrained ages range from 50$\pm$15\% to 79$\pm$02\%, while median relative uncertainties for the mass-constrained ages range from 42$\pm$4\% to 65$\pm$19\%.
However, we note that the decreases are not statistically significant in any case, i.e., the constrained and unconstrained medians agree within uncertainties for all model sets.
Meanwhile, the minimum relative uncertainties decrease in three cases, increase in one case \citep[non-magnetic]{Feiden2016}, and are unchanged in one case (PARSEC v1.2S).
The maximum relative uncertainty decreases in four cases and is unchanged in one (PARSEC v1.2S).
Taken together, these results suggest that our method of constraining the isochronal ages with dynamical mass likely improves the precisions of the derived stellar ages or, at the very least, does not degrade them.

\subsubsection{Increasing Agreement Among Model Sets}
When the unconstrained ages are used, the median age for Upper Sco ranges from 3 to 14~Myr, and the youngest and oldest ages do not agree within uncertainties.
In contrast, the median mass-constrained ages span a narrower range (6 to 12~Myr), agree within uncertainties in all cases, and in fact are identical for three of the five model sets (9~Myr).
This suggests that applying an independent mass constraint increases agreement between the model sets at the population level.

To test whether this phenomenon holds at the individual level, we examine the spread in age for each source for both the unconstrained and mass-constrained conditions.
We define the spread in age as $\sigma_\tau$ $=$ $\tau_{i,\,max}$ $-$ $\tau_{i,\,min}$, i.e. the difference between the highest and the lowest age returned by the model sets for a given source $i$.
A decrease in age spread indicates better agreement between the model sets for a given source.

Across the full sample, the minimum age spread drops from 5~Myr to 2~Myr and the maximum age spread drops from 63~Myr to 44~Myr.
The median age spread decreases from 11$\pm$6~Myr to 8$\pm$7~Myr.
The slight increase in scaled MAD is, somewhat ironically, driven by the overall decrease in $\sigma_\tau$: the mass-constrained $\sigma_\tau$ have a larger population of low values (5 sources with $\sigma_\tau$ less than half the median value) than the unconstrained $\sigma_\tau$ (2 sources with $\sigma_\tau$ less than half the median value).
This results in a slight increase in scaled MAD.

At the individual source level, we find a decrease in $\sigma_\tau$ for 14 out of the 17 sources that have at least two mass-constrained ages.
Of the remaining sources, two show an increase in $\sigma_\tau$ (J16090075$-$1908526, J16104636$-$1840598) and one shows no change (J16020757$-$2257467). 
For J16090075$-$1908526, the increase in $\sigma_\tau$ is driven by the \citet{Feiden2016} magnetic model. 
For J16104636$-$1840598, the increase is driven by the PARSEC v1.2S model. 
That these discrepant ages are driven by the magnetic \citet{Feiden2016} and PARSEC v1.2S models is consistent with their generally returning older ages than the other models. 
For each source, the age discrepancies of these models are present in both the unconstrained and mass-constrained cases, but are more pronounced in the mass-constrained case.

Because $\sigma_\tau$ is calculated from (at most) five ages, the presence of a single outlier can easily drive up its value. 
We recalculate $\sigma_\tau$ for these two sources excluding the relevant outlier model set under both constraint conditions.
We find that, when the outlier model set is excluded, the mass-constrained $\sigma_\tau$ is lower than the unconstrained $\sigma_\tau$ in both cases.
In other words, there is a very strong tendency for the dynamical mass constraint to produce increased agreement among the model sets, even in the presence of outliers.
When outlier ages do exist, the discrepancy appears to be confined to a single model set, and the remaining mass-constrained model sets are still in increased agreement compared to the unconstrained condition.

\subsection{Is Model Accuracy Correlated with Stellar Mass and/or Age?}
\label{age_mass_trends}
We found in \citet{Towner2025} that the PARSEC v1.2S models show a statistically-significant correlation between the isochronal-to-dynamical mass ratio and dynamical mass, i.e., they overestimate dynamical mass more severely at lower masses and underestimate it at higher masses.
Previous studies of isochronal mass and age have likewise found evidence of such variations \citep[e.g.][]{Rizzuto2016}, 
including some evidence of non-linear variations with mass \citep{Hillenbrand2004,David2019}. 
Here, we examine stellar ages and stellar masses together to search for any additional such trends. 

\subsubsection{Linear Correlations with Mass and Age}
\begin{figure*}
    \centering
    \includegraphics[width=\textwidth]{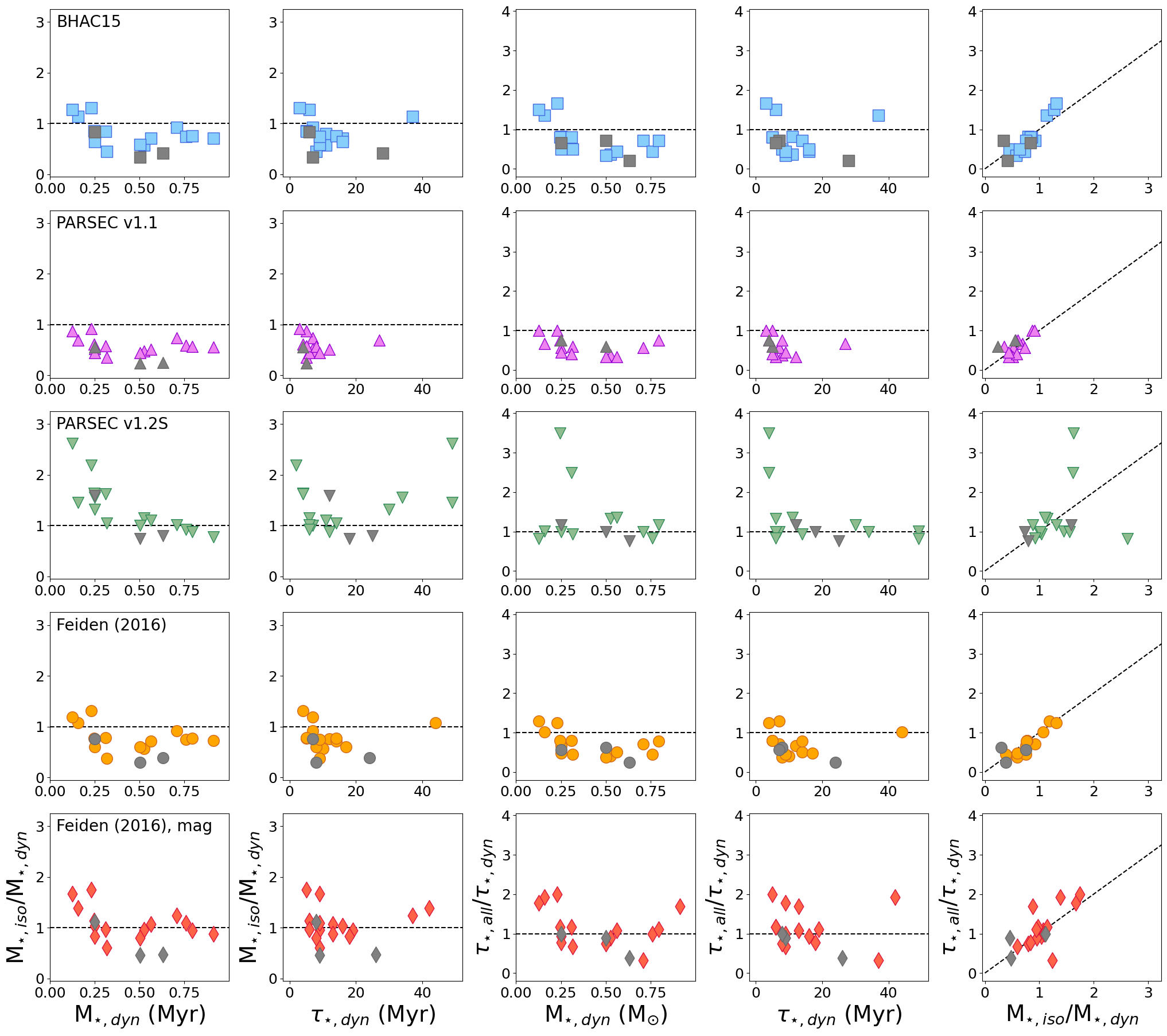}
    \caption{Trends in stellar age ratio and mass ratio with dynamical mass, mass-constrained age, and mass ratio. 
    From top to bottom, each row shows results for the BHAC15, PARSEC v1.1, PARSEC v1.2S, nonmagnetic, and magnetic \citet{Feiden2016} model sets. Sources with known or candidate companions are grey in all panels, but retain the symbol shape associated with each model set. From left to right, each column shows: mass ratio versus dynamical mass, mass ratio versus mass-constrained age, age ratio versus dynamical mass, age ratio versus mass-constrained age, and age ratio versus mass ratio. In all but the right-hand column, a ratio of 1:1 in the y-axis variable is indicated by a black, dashed horizontal line. In the right-hand column, the black dashed line represents a 1:1 (linear) relationship between age ratio and mass ratio. {\it Note: for source J16143367$-$1900133, the PARSEC v1.2S  models have an age ratio of $\tau_{\rm all}$/$\tau_{\rm dyn}$ = 7. For visual clarity, the y-axes of the third, fourth, and fifth columns only extend to $\tau_{\rm all}$/$\tau_{\rm dyn}$ $=$ 4, so this point is not visible. Were it visible, this source would appear in the middle row at (0.23,7) in column 3, (2,7) in column 4, and (2.2,7) in column 5.}
    }
    \label{age_mass_figure}
\end{figure*}

We first test for linear correlations between stellar mass, stellar age, and model accuracy. 
This analysis makes use of the ages presented in Table~\ref{ages_table}, the dynamical masses listed in Table~\ref{source_properties}, and the isochronal masses found in \citet[][their Table~6]{Towner2025}.
We calculate Spearman's $\rho$ correlation coefficients and $p$-values for each model set for the following quantities: isochronal-to-dynamical mass ratio versus dynamical mass, age ratio versus mass, mass ratio versus age, age ratio versus age, and mass ratio versus age ratio. 
We list the $\rho$ and $p$ values in Table~\ref{ratio_correlations}, and show scatterplots for each comparison in Figure~\ref{age_mass_figure}.
We repeat these calculations excluding those sources with confirmed or candidate companions, in case any trends are significantly influenced by those few sources.

We find that, in general, few model sets show statistically-significant correlations ($\geq$4$\sigma$) between dynamical mass or age and over- or underestimation of those values, but several do show $>$3$\sigma$ correlations. 
The PARSEC v1.2S models show a $>$4$\sigma$ correlation between mass ratio and mass, as noted in \citet{Towner2025}; this correlation is present regardless of whether sources with candidate companions are excluded.
The nonmagnetic \citet{Feiden2016} models show a $>$4$\sigma$ correlation between age ratio and mass ratio, i.e., the more severely those models underestimate mass, the more severely they underestimate age. 
This correlation strengthens when sources with companions are excluded, but remains between 4$\sigma$ and 5$\sigma$. 
The BHAC15 models show a $>$3$\sigma$ correlation between age ratio and mass ratio when all sources are considered, and this rises to $>$4$\sigma$ when potential-binary systems are excluded.

\begin{deluxetable*}{ll|cc|cc|cc|cc|cc}
\tablecaption{Spearman $\rho$ Correlation Coefficients for Isochronal versus Dynamical Stellar Mass and Age}
\tablecolumns{12}
\tablewidth{\textwidth}
\tablehead{
 \colhead{} & \colhead{} & \multicolumn{2}{c}{BHAC15} & \multicolumn{2}{c}{PARSEC v1.1} & \multicolumn{2}{c}{PARSEC v1.2S} & \multicolumn{2}{c}{F16, non-mag} & \multicolumn{2}{c}{F16, mag} \\
\colhead{x$_1$} & \colhead{x$_2$} & \colhead{$\rho$} & \colhead{$p^a$} & \colhead{$\rho$} & \colhead{$p$} & \colhead{$\rho$} & \colhead{$p$} & \colhead{$\rho$} & \colhead{$p$} & \colhead{$\rho$} & \colhead{$p$}
}
\startdata
\cutinhead{All Sources with $M_{\rm dyn}$ $\leq$ 1~\msun$^b$}
$M_{\rm dyn}$ & $M_{\rm iso}$/$M_{\rm dyn}$ & $-$0.51 & 0.03 & $-$0.31 & 0.2 & $-$0.84 & 1$\times$10$^{-5}$ & $-$0.44 & 0.07 & $-$0.49 & 0.04 \\
$\tau_{\rm dyn}$ & $M_{\rm iso}$/$M_{\rm dyn}$ & $-$0.44 & 0.08 & $-$0.23 & 0.4 & $-$0.11 & 0.7 & $-$0.43 & 0.09 & $-$0.14 & 0.6 \\
$M_{\rm dyn}$ & $\tau_{\rm all}$/$\tau_{\rm dyn}$ & $-$0.67 & 3$\times$10$^{-3}$ & $-$0.55 & 0.03 & $-$0.23 & 0.4 & $-$0.59 & 0.01 & $-$0.41 & 0.09 \\
$\tau_{\rm dyn}$ & $\tau_{\rm all}$/$\tau_{\rm dyn}$ & $-$0.47 & 0.05 & $-$0.53 & 0.04 & $-$0.58 & 0.01 & $-$0.43 & 0.08 & $-$0.29 & 0.2 \\
$M_{\rm iso}$/$M_{\rm dyn}$ & $\tau_{\rm all}$/$\tau_{\rm dyn}$ & 0.80 & 1$\times$10$^{-4}$ & 0.56 & 0.03 & 0.47 & 0.06 & 0.85 & 1$\times$10$^{-5}$ & 0.61 & 7$\times$10$^{-3}$ \\
\cutinhead{Excluding Sources with Potential Companions}
$M_{\rm dyn}$ & $M_{\rm iso}$/$M_{\rm dyn}$ & $-$0.51 & 0.05 & $-$0.33 & 0.2 & $-$0.91 & 3$\times$10$^{-6}$ & $-$0.46 & 0.09 & $-$0.47 & 0.07 \\
$\tau_{\rm dyn}$ & $M_{\rm iso}$/$M_{\rm dyn}$ & $-$0.41 & 0.1 & $-$0.33 & 0.3 & $-$0.02 & 0.9 & $-$0.43 & 0.1 & $-$0.05 & 0.8 \\
$M_{\rm dyn}$ & $\tau_{\rm all}$/$\tau_{\rm dyn}$ & $-$0.71 & 5$\times$10$^{-3}$ & $-$0.54 & 0.06 & $-$0.11 & 0.7 & $-$0.61 & 0.02 & $-$0.41 & 0.1 \\
$\tau_{\rm dyn}$ & $\tau_{\rm all}$/$\tau_{\rm dyn}$ & $-$0.38 & 0.2 & $-$0.45 & 0.1 & $-$0.60 & 0.02 & $-$0.31 & 0.3 & $-$0.21 & 0.4 \\
$M_{\rm iso}$/$M_{\rm dyn}$ & $\tau_{\rm all}$/$\tau_{\rm dyn}$ & 0.88 & 4$\times$10$^{-5}$ & 0.65 & 0.02 & 0.34 & 0.2 & 0.91 & 6$\times$10$^{-6}$ & 0.58 & 0.02
\enddata
\tablenotetext{a}{For normally-distributed data, p-values correspond to $\sigma$ values according to: 1$\sigma$: p $\leq$ 0.68; 2$\sigma$: p $\leq$ 0.05; 3$\sigma$: p $\leq$ 2.7$\times$10$^{-3}$; 4$\sigma$: p $\leq$ 6.3$\times$10$^{-5}$; 5$\sigma$:  p $\leq$ 5.7$\times$10$^{-7}$.} 
\tablenotetext{b}{In practice, this mass restriction only impacts the $M_{\rm dyn}$ versus $M_{\rm iso}$/$M_{\rm dyn}$ correlations. None of the sources with $M_{\rm dyn}$ $>$ 1~\msun\/ have dynamical mass-constrained ages in Table~\ref{ages_table}, so any correlations involving $\tau_{\rm dyn}$ exclude these sources automatically. $M_{\rm dyn}$ versus $M_{\rm iso}$/$M_{\rm dyn}$ correlation statistics for the full sample can be found in \citet{Towner2025}, Table 7.}
\label{ratio_correlations}
\end{deluxetable*}

However, the Spearman $\rho$ correlation test (along with similar tests such as Pearson, Kendall's $\tau$, etc) only capture linear correlations and not more complex trends such as those noted in \citet{David2019}.
Therefore, in addition to the statistical analysis discussed above, we examine our distributions directly to search for more complex trends. 

\subsubsection{Non-linear Trends with Mass and Age}
Figure~\ref{age_mass_figure} shows scatterplots of the isochronal/dynamical mass ratios and age ratios plotted against dynamical mass, against mass-constrained age, and against each other.
With the exception of the PARSEC v1.2S models, which show a generally linear negative correlation between mass ratio and mass, most models show non-linear behavior.
In general, the most severe underestimation of mass occurs in the $\sim$0.25 to $\sim$0.6~\msun\/ range, with improvements at both lower and higher masses (see the left-hand column of Figure~\ref{age_mass_figure}).
This non-linear, mass-dependent behavior is broadly consistent with the trends noted by \citet{David2019}.
Those authors measure stellar mass for eclipsing binaries and estimate stellar ages using both the mass-radius diagram (MRD) and Hertzsprung-Russell diagram (HRD).
To derive stellar ages, they use the five model sets we consider here, along with the Mesa Isochrones and Stellar Tracks models \citep[MIST,][]{Choi2016,Dotter2016}, the PARSEC v1.0 models of \citet{Girardi2000}, and both a 50\% and a 0\% starspot-coverage fraction in the SP15 models \citep{Somers2015}.
They find that the discrepancies between the isochronal masses and their MRD/HRD-derived masses gradually worsen below 1~\msun, with a maximum discrepancy at $\sim$0.3~\msun\/ and improvement toward 0.1~\msun.

We note that, despite their overall 1:1 correlation with dynamical mass in \citet{Towner2025} for 0.1~\msun\/ $\leq$ \mstar\/ $\leq$ 1~\msun, the magnetic \citet{Feiden2016} models are not exempt from these non-linear trends; indeed, they show a maximum overestimation of mass up to $\sim$100\% in the $\leq$0.25~\msun\/ mass range, and maximum underestimation of $\sim$50\% at $\sim$0.3~\msun\/ (excluding the confirmed or candidate binary sources marked in grey). 
We thus revise our assessment of these models from \citet{Towner2025}: although we still find no evidence of systematic overestimation for \mstar\/ $\leq$ 0.6~\msun, we do find find a tendency toward overestimation in the very low mass range of \mstar\/ $\leq$0.25~\msun. 

When we examine age ratio versus dynamical mass (middle column), we find that similar patterns exist for most model sets (again excepting the PARSEC v1.2S models): the models are most likely to underestimate age for stars in the $\sim$0.25 to 0.75~\msun\/ range, with some improvement toward higher masses.
The most severe underestimation of age appears to occur at \mstar\/ $\approx$ 0.5~\msun.
The greatest likelihood of {\it over}estimating age occurs for the lowest-mass (\mstar\/ $\lesssim$0.25~\msun) sources, but only for the \citet{Feiden2016} model sets (both magnetic and non-magnetic) and the BHAC15 models.
The PARSEC v1.1 models never overestimate age, just as they never overestimate mass.
The PARSEC v1.2S models show no clear trends with mass, except that they seem to overestimate age most severely for \mstar\/ $\leq$ 0.5~\msun.

The fact that the age ratios mirror the behavior of the mass ratios can be explained by the right-hand column of Figure~\ref{age_mass_figure}.
This column shows that these two ratios have a more or less linear relationship for the BHAC15, PARSEC v1.1, and nonmagnetic \citet{Feiden2016} models. 
The relationship is also broadly linear for the magnetic \citet{Feiden2016} models, but with notably more scatter than the previous three model sets.
This can explain the overall similarities in the trends between the leftmost and middle columns in Figure~\ref{age_mass_figure} for all but these four model sets.
The PARSEC v1.2S models show no clear relationship between age ratio and mass ratio, which is consistent with the lack of similarity between those two distributions in the middle row. 

We do not find any strong trends (statistically or visually) between mass ratio and age, or between age ratio and age.
In other words, the strongest trends in model age and mass accuracy appear to be mass-dependent, not age-dependent, for our sample.

\subsubsection{On the Consistency of Trends Across Model Sets}
The consistent presence, and especially the consistent {\it shape}, of these non-linear trends across most PMS model sets is striking.
It is beyond the scope of this paper to robustly determine the origin of this trend, but we briefly discuss two possibilities here.

The first possibility is that there is some systematic inaccuracy in the input data.
In this work, we use the same $T_{\rm eff}$ and \lstar\/ across all model sets for a given source.
The $T_{\rm eff}$ are converted from the spectral types of \citet{Luhman2012} by \citet{Barenfeld2016}, who use the temperature scales of \citet{Schmidt-Kaler1982}, \citet{Straizys1992}, and \citet{Luhman1999}.
The \lstar\/ are derived by fitting BT-Settl atmospheric models \citep{Allard2011} to optical and near-infrared photometric data and holding $T_{\rm eff}$ fixed \citep[see][Appendix A for further details of the SED fitting procedure]{Towner2025}.
The $T_{\rm eff}$ and \lstar\/ of \citet{David2019}, meanwhile, are determined by SED fitting to BT-Settl atmospheric models in which $T_{\rm eff}$ and stellar radius (among other free parameters) are allowed to vary, and \lstar\/ is calculated using the Stefan-Boltzmann equation. 
If there is a systematic inaccuracy in the input data, it is not immediately clear why it would be present in both works, or whether it is present in the original photometry, the spectral type identifications, conversion from spectral type to $T_{\rm eff}$ (where applicable), or derivation of \lstar.

The second possibility is that these model sets, regardless of their population-level accuracy, may be missing some relevant stellar physics in the 0.1 $\leq$ \mstar\/ $\leq$ 1~\msun\/ range.
If so, all four model sets must be missing the same physics to produce such a consistent trend: while the introduction of a global magnetic field in the \citet{Feiden2016} magnetic models does appear to have produced a population-level improvement in mass and age accuracy, the mass-dependent trends still remain. 
Because these trends are also present in \citet{David2019}, any ``missing physics'' issue would have to encompass the additional model sets examined in that work as well (MIST, PARSEC v1.0, and the SP15 models).

One phenomenon that potentially satisfies both possibilities is starspots. 
Starspots can alter both the photometric and spectral properties of a given star, and their core properties (size, umbral and penumbral temperatures, magnetic field strengths) and surface filling fractions appear to vary with stellar mass, temperature, and age \citep[see review by][and references therein]{Berdyugina2005}.
This can lead to mass- and age-dependent shifts in observed colors, including nonlinear shifts \citep[bluer at shorter wavelengths, redder at longer wavelengths; see][]{Somers2015}.
Starspots are also a natural consequence of global magnetic fields such as those of \citet{Feiden2016}, but are not explicitly included in those models; this could explain why the \citet{Feiden2016} magnetic models show a population-level improvement over standard models, yet still exhibit mass-dependent trends in Figure~\ref{age_mass_figure}.
A model that combines (global) magnetic fields with the observational color and temperature effects of (localized) starspots, then, could potentially be a fruitful avenue of future research. 
However, as we do not test any starspot models in this work, such a possibility is highly speculative at this time.

\subsection{Overall Evaluation of the PMS Model Sets}
\label{model_evaluation}
Based on our findings in the previous sections, and our results in \citet{Towner2025}, we conclude that the magnetic models of \citet{Feiden2016} are the most reliable isochronal method (of the five examined in this work) for determining stellar mass and age in the absence of independent mass constraints.
The global magnetic fields in these models are thought to inhibit convection in low-mass stars, which causes them to contract more slowly along the Hayashi track than is predicted by standard models \citep[][and references therein]{Feiden2016}.
When placed on an HR diagram, these models predict older ages and larger masses for a given \lstar\/ and $T_{\rm eff}$ than nonmagnetic models.
These \mstar\/ and $\tau$ are in better agreement with our dynamical results than predictions from models that do not explicitly include magnetic fields.

However, we caution that these magnetic models can have high deviations from the dynamical mass-constrained results for individual sources.
They are also subject to the same mass-dependent mass and age accuracy issues present in the BHAC15, PARSEC v1.1, and nonmagnetic \citet{Feiden2016} models.
Any isochronal method should be used with caution for individual sources, and isochronal results are most reliable when used for large populations. 

We also note that, although the PARSEC v1.2S models do have a median unconstrained-versus-constrained age ratio of 1 and median $\Delta\tau$ $=$ 0, we do not consider them to be as reliable as the magnetic models of \citet{Feiden2016}.
They have a statistically-significant tendency to overestimate mass, a consistently higher scaled MAD for median population age than the other model sets, and exhibit greater change in median age when constrained compared to the magnetic \citet{Feiden2016} models.
Their age accuracy does not appear to be consistent across the sample, nor are their deviations in age systematic as far as we can tell.
This makes them unpredictable for any individual source and somewhat unstable even at the population level.
It is beyond the scope of this paper to examine why the PARSEC v1.2S models behave so differently from the other four model sets tested.
However, we do conclude that, for the mass and age range covered by our sample, the PARSEC v1.2S models are not as reliable as the other four model sets tested.

\section{Comparison with Existing Literature for Upper Sco}
\label{age_mass_literature}
Numerous methods exist for deriving stellar mass and age: orbital dynamics in binary systems, dynamical traceback modeling, and the lithium-depletion boundary (LDB), to name a few \citep[see e.g.][and references therein]{Soderblom2014,Manara2023_review}.
In this section, we compare our results derived from mass-constrained isochrone fitting to selected results for Upper Sco using these alternate methods. 
However, some methods are known to be less accurate for young and/or low-mass sources \citep[e.g. lithium depletion boundary for $\tau$ $\lesssim$ 20~Myr, lithium abundances for \mstar\/ $\lesssim$ 1~\msun\/ and age $\lesssim$ 50~Myr; see][and references therein]{Soderblom2014}.
As Upper Sco is generally considered to be  younger than this (5-11~Myr), we restrict our comparison to two alternate methods: eclipsing binaries and dynamical traceback modeling. 
In both cases, we compare our results to studies performed for Upper Sco specifically.

\subsection{Mass and Age from Orbital Dynamics}
\citet{Rizzuto2016} use astrometric measurements to derive the orbital and stellar properties of seven G-, K-, and M-type binary pairs in Upper Sco.
They also derive stellar mass and age using three PMS evolutionary models: the Padova models of  \citet{Girardi2002}, the Dartmouth models of \citet{Dotter2008}, and the BT-Settl atmospheric models of \citet{Allard2011}. 
They find that, for the G-type stars in their sample, their isochronal and dynamical results agree within uncertainties for both mass and age.
However, 
all three models produce younger ages for the M-type stars ($\sim$7~Myr) than for the G-type stars ($\sim$11.5~Myr).
These M-star ages do not agree within uncertainties with the G-type ages in that work.
The authors suggest that this may indicate calibration issues in the stellar evolutionary models for the subsolar regime. 

Likewise, \citet{David2019} examine nine eclipsing binary systems in Upper Sco ranging in mass from 0.11~\msun\/ to 5.58~\msun.
There is no overlap between this sample and the sources examined in \citet{Rizzuto2016}.
\citet{David2019} derive dynamical masses for their targets and compare these to isochronal predictions for mass and age using both Hertzsprung-Russell Diagrams (HRDs) and Mass-Radius Diagrams (MRDs).
In addition to the mass-dependent trends discussed in \S~\ref{age_mass_trends} above, \citet{David2019} find that standard PMS models and the SP15 spot-free models produce a consistent, median age of 5-7~Myr for Upper Sco using HRDs, while the \citet{Feiden2016} magnetic and SP15 50\%-spotted models produce median ages of 9-10~Myr instead. 

These findings are consistent with ours: standard PMS evolutionary models tend to produce younger ages than those that account for the effects of magnetic fields, and models that account for magnetic fields produce ages of $\sim$10~Myr for Upper Sco rather than the younger $\sim$5~Myr ages commonly derived from low-mass stars.

\subsection{Age from Stellar Proper Motions and Dynamical Traceback Modeling}
Dynamical traceback modeling uses the observed proper motion of stars in young clusters in combination with N-body modeling to derive the cluster age.
\citet{Miret-Roig2022} used proper-motion data and dynamical traceback modeling of 2190 sources in Upper Scorpius and Ophiuchus to determine association ages. 
Based on the sources' distributions in 6D phase space and a Gaussian Mixture Model (GMM) analysis, they identify 7 subgroup within Upper Sco and Ophiuchus.
Among the four subgroups belonging to Upper Sco, traceback ages range from 0.3$\pm$0.5~Myr to 4.6$\pm$1.1~Myr, with a median of 2.3~Myr.
They define ``traceback age'' in this context to mean the time at which the cluster members occupied the minimum total volume in space. 

\citet{Miret-Roig2022} report that the typical traceback ages for Upper Sco are $\lesssim$5~Myr, which is 1-3~Myr younger than the isochronal ages they derive using the PARSEC isochrones of \citet{Marigo2017}. 
Their results are even more discrepant with the mass-constrained ages we derive in this work. 
This discrepancy is likely attributable to the difference in age derivation method.
\citet{Miret-Roig2022} note that their model does not account for perturbations in stellar orbits due to interactions between group members, dissipation of the parent cluster gas, and stellar feedback.
They suggest that including these effects in the model could account for the discrepancy between their isochronal and dynamical traceback ages.
Studies of other nearby OB associations \citep[e.g. TW Hya and $\beta$ Pic; see][and references therein]{Soderblom2014} have also found a tendency of dynamical traceback modeling to predict younger group ages compared to other methods.

\section{Testing for Age Bias in Our Sample}
\label{age_bias}
In this section, we test whether our sample - and, by extension, the median age we derive for our sources - is representative of Upper Sco as a whole. 
Given our conclusions in \S~\ref{model_evaluation}, we use the magnetic \citet{Feiden2016} models to conduct this evaluation. 

We compare our results to the much larger Sco-Cen sample of \citet{Fang2025}. 
Those authors use {\it Gaia} distances and spectroscopy, combined with archival photometry, to derive stellar luminosities, spectral types and temperatures, and A$_V$ for 8,846 sources in the full Sco-Cen region. 
We retrieve stellar luminosities and temperatures for all K- and M-type stars from \citet[][their Table~1]{Fang2025}.
We use the `SpT2' column for spectral type classifications when available \citep[the classifications determined directly by][]{Fang2025}; when no such classification is available, we use the literature spectral types listed in the `SpT1' column. 

We combine these data with the Sco-Cen kinematic subgroup classifications of both \citet{Luhman2022} and \citet{Ratzenbock2023a} to identify the K- and M-type members of the Upper Sco spatio-kinematic subgroup(s) specifically. 
Using the \citet{Luhman2022} group classifications, we consider all K- and M-type stars that are primarily associated with Upper Sco (kinematic groups `u,' `ul,' `ulc,' and `uc' in that system).
We find a total of 2059 sources using these criteria.
Using the \citet{Ratzenbock2023a} group classifications, we consider all K- and M-type stars in SigMA groups 1 through 9 inclusive \citep[kinematic groups listed as belonging to Upper Sco in Table~3 of][]{Ratzenbock2023a}. 
We find a total of 2469 sources using these criteria.

We generate isochronal masses and ages for these large samples using the \citet{Feiden2016} magnetic models.
Both the Luhman-selected and Ratzenb\"{o}ck-selected samples have $>$100 sources with \mstar\/ between 0.08499 and 0.08501~\msun, i.e., they appear to contain a ``pile-up'' of sources at the low-mass edge of the \citet{Feiden2016} magnetic model grid (122 and 147 sources, respectively).
We calculate all statistics in the following discussion both with and without these grid-edge sources.

\begin{figure*}
    \centering
    \includegraphics[width=0.49\textwidth]{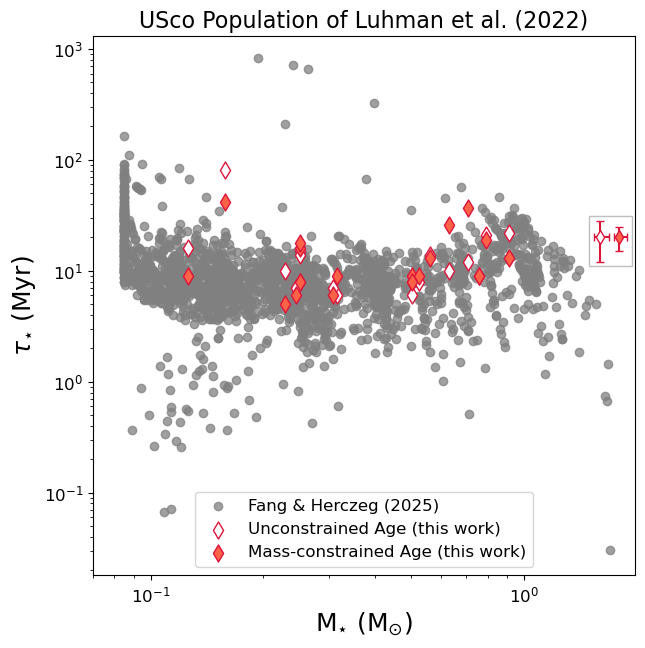}
    \includegraphics[width=0.49\textwidth]{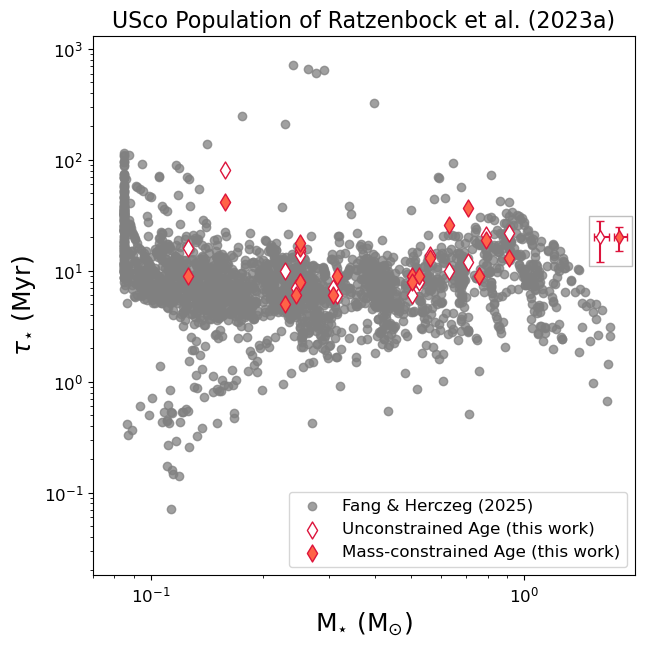}
    \caption{Stellar age versus mass for our unconstrained (open diamonds) and mass-constrained (solid diamonds) sample using the magnetic models of \citet{Feiden2016}. Typical (median) error bars for the two cases are denoted by the boxed symbols on the right-hand side of each panel. Larger samples of K- and M-type members from Upper Sco are shown as solid gray circles in both panels. Membership in Upper Sco is determined using the membership classifications of ({\it left}) \citet{Luhman2022} and ({\it right}) \citet{Ratzenbock2023a}. Stellar masses and ages for the larger samples are calculated using the \lstar\/ and $T_{\rm eff}$ of \citet{Fang2025} and the magnetic models of \citet{Feiden2016}. There is a clear ``pile-up'' of sources at $\approx$0.085~\msun, which is the low-mass edge of the magnetic \citet{Feiden2016} model grid. We calculate KS and Anderson-Darling statistics for our sample versus both populations, and find no statistically-significant difference between our sample and the larger Upper Sco samples shown here.}
    \label{age_bias_test}
\end{figure*}

Figure~\ref{age_bias_test} shows the magnetic \citet{Feiden2016} age versus mass for the full K- and M-type stellar population in Upper Sco from \citet{Fang2025}.
We show results using \citet{Luhman2022} selection criteria in the left-hand panel, and results using the \citet{Ratzenbock2023a} selection criteria in the right-hand panel.
The dynamical masses for our sample, and the unconstrained and mass-constrained ages, are overlaid.

The median age of Luhman-selected sample of K- and M-stars is 8$\pm$4~Myr both with and without the ``pile-up'' sources.
The median age of the Ratzenb\"{o}ck-selected sample is 8$\pm$4~Myr when the grid-edge sources are included, and 7$\pm$4~Myr when they are excluded.
These values are slightly lower than, but still in statistical agreement with, the median unconstrained and mass-constrained ages for our dynamical-mass sample of 10$\pm$5 and 9$\pm$5~Myr, respectively.

We perform a 2-sample Kolmogorov-Smirnov test (KS test) to evaluate the statistical likelihood that our dynamical mass sample is drawn from a different parent population than Upper Sco as a whole. 
Regardless of which Upper Sco selection criteria are used, and regardless of the inclusion/exclusion of the grid-edge sources, we find no statistically-significant difference between our dynamical-mass sample and the broader Upper Sco population. 
This is true for both our unconstrained and mass-constrained ages. 
The KS test statistic values range from 0.29 $<$ D $<$ 0.39, and $p$-values range from 0.006 to 0.09.
None of these $p$-values fall below the common statistical-significance cutoff of $p$ $<$ 0.001, much less the more stringent 4$\sigma$ cutoffs we use in this work.

We also perform a 2-sample Anderson-Darling test.
While the KS test is more sensitive to differences in population medians, the Anderson-Darling test is more sensitive to differences in the tails of population distributions.
We again find no statistically-significant difference between our sample and the Upper Sco population as a whole.
The Anderson-Darling test statistic values range from 2.06 to 5.43, and the $p$-values range from 0.002 to 0.05. 
The lower $p$-values of the Anderson-Darling test are consistent with greater sensitivity to outlier ages in the Upper Sco sample, but still do not rise to the level of statistical significance.
To the extent that there differences between our sample and the overall population of K- and M-type stars in Upper Sco, these may be attributable to our significantly smaller sample size, and in any case, the population medians all agree within their respective uncertainties.

We thus conclude that our derived $\sim$9~Myr age for Upper Sco using our dynamical-mass constrained, K- and M-type PMS stars is robust against sample age bias.

\section{Implications for Protoplanetary Disk Lifetimes}
\label{disk_lifetimes}
Protoplanetary disk masses have long been known to decrease with stellar age, due to a combination of inner-disk clearing by the host star and evolution of disk solids with time \citep[see review by][and references therein]{Manara2023_review}.
However, disks do not evolve uniformly; there is evidence that dust content in the inner disk region evolves more rapidly than dust in the outer disk \citep[e.g.][]{Ribas2014,Ben2025}.
Disk lifetimes and other properties (e.g. luminosity) are also known to vary with stellar mass, with older, more luminous disks preferentially being hosted by low-mass stars \citep{Carpenter2006,Pfalzner2024,Carpenter2025}.

Most calculations of disk lifetime do not come from the ages of individual protostellar/PMS systems, as individual systems are known to exhibit a wide range of ages \citep[$<$1 to $>$40~Myr, see][]{Pfalzner2024}.
Instead, typical disk lifetimes are calculated from the fraction of stars hosting disks in regions or subregions of different ages \citep[e.g.][]{Fang2025}.
The disk lifetime or half-life timescales therefore depend less on observations of individual disks than on accurate derivations of the ages of overall stellar populations. 

Many literature studies using disk fractions derived from near-infrared data have suggested disk half-life timescales of 1$\sim$4~Myr, depending on stellar mass, metallicity, and cluster environment \citep{Mamajek2009,Fedele2010,Fang2013}.
However, recent evidence has suggested that disk lifetimes may be significantly longer than this.
\citet{Ben2025} recently showed that disk fractions derived using 1.6$-$4.6~\mum\/ infrared excess yield typical disk lifetimes of 1.6$\pm$0.1~Myr, but fractions derived using 12~\mum\/ excess yield lifetimes of 4.4$\pm$0.3~Myr.
These results are consistent with those of \citet{Ribas2014}, who find disk decay timescales of 2$-$3~Myr using 3.4$-$12~\mum\/ data, but 4.2$-$5.8~Myr using 22$-$24~\mum\/ data.
\citet{Ribas2014} interpret this as the slower evolution of dust at larger radii within the disk.
Additionally, \citet{Fang2025} calculate a disk decay timescale for the Sco-Cen complex using infrared excess in the {\it Spitzer} IRAC or WISE W2/W3 bands, the Sco-Cen subgroups identified by \citet{Ratzenbock2023a}, and ages derived using the Stellar Parameters of Tracks with Starspots \citep[SPOTS;][]{Somers2020} models with a spot coverage fraction of 0.51.
They derive a disk decay timescale of $\tau$ $=$ 6.0$\pm$2.0~Myr, which is approximately double that found in previous works \citep[][and references therein]{Fang2025}.

Our results are consistent with such findings.
Using standard, nonmagnetic PMS evolutionary models, we find that median stellar age increases by a factor of $\sim$2 when a mass constraint is applied \citep[BHAC15, PARSEC v1.1, nonmagnetic][]{Feiden2016}.
Using magnetic models, we find a median age that is consistently $\sim$ 2-3$\times$ the unconstrained ages returned by standard models.
If such underestimations of stellar age are common across multiple star-forming regions, then a typical increase in disk lifetime of $\gtrsim$2$\times$ would not be unreasonable.
Such an increase would suggest at least an additional $\sim$4~Myr for planet formation to occur.

\section{Summary}
\label{summary}
We have derived the ages of 23 pre-main sequence K- and M-type stars with disks in Upper Scorpius using dynamical masses and pre-main sequence evolutionary models. 
We use five PMS model sets: BHAC15 \citep{Baraffe2015}, PARSEC v1.1 \citep{Bressan2012}, PARSEC v1.2S \citep{Chen2014}, and both the standard and magnetic models of \citet{Feiden2016}.
We derive stellar ages in both unconstrained and dynamical mass-constrained cases, and compare our findings under both conditions. 

Our overall evaluations of the five model sets examined in this work are as follows:

\begin{enumerate}
\item The magnetic models of \citet{Feiden2016} show the greatest agreement with our dynamical results for stellar mass and mass-constrained stellar age simultaneously. The magnetic model yields median stellar ages of 9-10~Myr for our sample regardless of mass constraint, which is consistent with the older ages for Upper Sco in the literature derived using higher-mass sources \citep{Pecaut2012}. 
These models show more scatter in $M_{\rm iso}$/$M_{\rm dyn}$ and $\tau_{\rm all}$/$\tau_{\rm dyn}$ than the BHAC15, PARSEC v1.1, and non-magnetic \citet{Feiden2016} models, but less than the PARSEC v1.2S models. 
\item The BHAC15, PARSEC v1.1, and nonmagnetic \citet{Feiden2016} models have a tendency to systematically underestimate both stellar mass and stellar age at the population level. They have comparatively tight correlations between $M_{\rm iso}$/$M_{\rm dyn}$ and $\tau_{\rm all}$/$\tau_{\rm dyn}$, i.e., relative inaccuracies in stellar mass are likely to yield comparable relative inaccuracies in stellar age. These model sets produce young ages (3-5~Myr) for Upper Sco when not independently constrained, but older ages (6-9~Myr) when constrained by a dynamical mass for each source. The PARSEC v1.1 models consistently yield the lowest masses and youngest ages for our sample.
\item The PARSEC v1.2S models have a strong tendency to overestimate stellar mass for \mstar\/ $\lesssim$ 0.5~\msun, and a mild tendency to underestimate stellar mass for \mstar\/ $\gtrsim$ 0.5~\msun. This trend is statistically significant at the $>$4$\sigma$ level. The PARSEC v1.2S models can overestimate stellar age by as much as $\sim$600\% for individual sources, and consistently return the oldest age for Upper Sco at the population level. 
These models yield a median stellar age of 12-14~Myr for Upper Sco, and consistently show the largest scaled MAD of all model sets. Given their somewhat unstable and non-systematic behavior, we consider this to be the least reliable model set of the five examined here.
\item Four of the five models (excepting PARSEC v1.2S) show the same mass-dependent behavior: the models are most likely to overestimate mass and age for \mstar\/ $\lesssim$ 0.25~\msun, they are most likely underestimate mass and age for \mstar\/ $\sim$0.5~\msun, and they show improved agreement with dynamical results as \mstar\/ rises to $\gtrsim$ 0.75~\msun. 
Similar mass-dependent trends have been observed in previous literature, but the source of such trends is unclear. We suggest that this may be a fruitful avenue for future research, as there may be a common underlying cause in either the PMS models or the input data (or both). 
\end{enumerate}

Applying an independent dynamical-mass constraint to the isochronal ages preferentially increases stellar age for three of the five model sets tested here.
Moreover, this constraint brings most model sets into closer alignment with the older, 8-11~Myr age long derived for Upper Sco using early-type sources, and appears to ease, if not resolve, the longstanding age discrepancy for this region.

Additionally, applying the independent mass constraint consistently brings the five model sets into much closer alignment with each other.
This is true at both the population level (median age for each model) and at the individual source level (age spread, $\sigma_{\tau}$).
We suggest that applying such an independent mass constraint, where available, can help to relieve some of the strong model-based differences in stellar age long observed in the literature. 

We compare our results to studies of Upper Sco that use alternate methods to determine age, and find agreement with results derived using eclipsing binaries \citep{Rizzuto2016,David2019}.
In contrast, our derived stellar ages are $\gtrsim$ 4~Myr older than those obtained from dynamical traceback methods using bulk stellar populations \citep{Miret-Roig2022}.
We suspect this is due to the difference in methods and the challenges of dynamical traceback modeling in particular $-$ specifically, the challenge in accounting for perturbations in stellar paths due to gravitational interactions with other group members (ibid). 

We test for age bias in our sample by comparing our results to those from a much larger sample of K- and M-type sources in Upper Sco.
Neither a KS test nor an Anderson-Darling test show any statistically-significant differences between our sample of 23 sources and the larger sample of low-mass sources from \citet{Fang2025}.
We conclude that our sample is not biased in age, and that our mass-constrained ages of $\sim$9~Myr are representative of the low-mass PMS population of Upper Sco as a whole.

We consider the implications of our findings for disk lifetimes and planet-formation timescales.
We find that the typical increase in median stellar age from standard models ($\sim$2$\times$ when the mass constraint is applied) is consistent with literature findings that suggest a median disk lifetime of $\sim$2$\times$ previous estimates \citep[e.g. 6~Myr compared to 2-3~Myr;][and references therein]{Fang2025}.
Our results suggest that planet-formation timescales may thus be $\gtrsim$4~Myr longer than previously thought.

\begin{acknowledgements}
We offer our sincere thanks to M. Fang and G. Herczeg, who were kind enough to share their stellar data for Sco-Cen prior to formal publication of those data on CDS. 
This paper makes use of the following ALMA data: ADS/JAO.ALMA\#2019.1.00493.S. 
ALMA is a partnership of ESO (representing its member states), NSF (USA) and NINS (Japan), together with NRC (Canada), NSTC and ASIAA (Taiwan), and KASI (Republic of Korea), in cooperation with the Republic of Chile. 
The Joint ALMA Observatory is operated by ESO, AUI/NRAO and NAOJ.
The National Radio Astronomy Observatory is a facility of the National Science Foundation operated under cooperative agreement by Associated Universities, Inc.
The results reported herein benefitted from collaborations and/or information exchange within NASA's Nexus for Exoplanet System Science (NExSS) research coordination network sponsored by NASA's Science Mission Directorate under Agreement No. 80NSSC21K0593 for the program ``Alien Earths''.'' 
\end{acknowledgements}

\software{
astropy \citep{astropy_i,astropy_ii,astropy_iii}, 
pdspy \citep{Sheehan2019}
}

\end{document}